# Multilateral Clearing on Invoice Graphs: Path Enabled Compensation Versus Cycle Restricted Netting

Peplluis de la Rosa Esteva[ab], Sai Srikanth Madugula[a]

[a] Woxsen University, Hyderabad, India

[b] Universitat de Girona, Catalonia, Spain, EU

Corresponding author: Peplluis de la Rosa Esteva[12] Josepllluis.delarosa@udg.edu

**Abstract**

Late payments and limited working capital propagate liquidity stress across supply chains, especially among small and medium sized enterprises. This paper develops a path enabled clearing framework for invoice backed trade networks and evaluates Byppay, a local compensation procedure based on repeated three party reductions. Unlike cycle restricted netting, which can clear only obligations embedded in directed cycles, Byppay can also reduce obligations along open chains while preserving node net positions on a combined post settlement state composed of the residual invoice graph and a generated settlement instruction layer. The paper contributes by formalizing this framework for invoice networks, defining a comparable post settlement metric system, specifying deterministic benchmark procedures, and assessing them on the same real invoice graphs. The empirical analysis uses the 2021 IMI invoice corpus of 133,191 invoices totaling EUR 19.67 billion. On the annual aggregate, the cycle restricted Netting benchmark achieves EUR 4.13 billion of debt relief, or 20.99%, whereas Byppay achieves EUR 10.60 billion, or 53.87%. In the metric system adopted here, gross obligation reduction and liquidity relief are both measured on post settlement payable mass and are therefore identical for both procedures. The results show that path enabled local compensation reaches a substantially larger reduction fixed point than a cycle restricted benchmark on the same invoice graphs, while also supporting a deployable execution logic compatible with selective disclosure and distributed coordination.

## 1 Problem and Contribution

This paper compares two settlement move sets on the same invoice graphs. The first is cycle restricted netting, represented by a deterministic benchmark implementation of the Romanian Netting logic described by Gavrilă and Popa (2021). The second is Byppay, a path enabled local procedure that repeatedly applies a three party reduction on a directed path A to B to C. The structural distinction is simple but consequential. Cycle based methods can act only on obligations embedded in directed cycles, whereas a local path move can also act on open chains. In supply chain invoice graphs, where feed forward and weakly reciprocal structures are common, that difference can materially affect the reduction fixed point reached by the procedure.

This paper introduces a path enabled clearing framework for invoice backed trade networks in which local three-party reductions can operate on both open chains and cycles. The contribution lies in integrating this move set into an invoice network model, defining a comparable post settlement metric system, benchmarking it deterministically against an operational cycle restricted baseline on the same graphs, and specifying an execution logic compatible with selective disclosure and distributed coordination. The paper adopts one interpretation consistently. Byppay generates a settlement instruction layer on top of the invoice graph. It does not automatically novate or replace the underlying invoice claims. The residual invoice graph continues to represent the original claims, while the instruction layer records how settlement is redirected after local path compression. This framing permits a precise comparison with cycle based netting and avoids overclaiming legal effects that are not established by the benchmark itself.

The paper makes four contributions. First, it formalizes the invoice graph, the residual graph, the generated settlement instruction layer, and the performance metrics required to compare path enabled and cycle restricted procedures on a common economic basis. Second, it specifies exact deterministic pseudocode for Byppay and for the cycle based Netting benchmark used in the empirical study. Third, it reports a like for like benchmark on the 2021 IMI invoice corpus, using identical monthly graphs and method specific fixed-point stopping conditions. Fourth, it outlines a prototype execution architecture for local Byppay operation on a permissioned ledger.

Taken together, the contribution of this work is best understood as a deployable protocol supported by empirical evidence. The remainder of the paper is organised as follows. Section 2 reviews related work on debt clearing and privacy aware settlement. Section 3 introduces the problem setting, the invoice graph model, the Byppay local path step, and the associated notation and metrics. Section 4 develops the temporal and topological analysis, including theorems, propositions, and illustrative examples that clarify the structural advantage of path enabled compensation over cycle restricted netting. Section 5 presents the benchmark design, describes the invoice corpus, and reports the empirical results for both a representative month and the full year. Section 6 discusses deployment implications, including smart contract execution, concurrency safeguards, selective disclosure, and legal enforceability. Section 7 concludes by summarising the findings, discussing the counterparty risk trade off introduced by path-based moves, and outlining directions for future research.

## 2 Related work: Debt Clearing

The literature on multilateral debt clearing spans three partially overlapping strands. A first strand studies debt simplification as a graph transformation that preserves node net positions. Verhoeff (2004) gives a clear formulation of the multiple debt settlement problem and discusses algorithmic variants in which obligations are compressed without changing final net positions. The Byppay step used here belongs to this broad family of local debt simplification primitives.

A second strand studies global optimization under settlement constraints. Abraham, et al. (2007) show that constrained barter and clearing variants can become computationally hard when the admissible settlement structure is restricted. This literature is important because it clarifies the distinction between a globally optimal clearing plan and a practical greedy reduction procedure. The present paper does not solve a global optimization problem. It compares two deterministic reduction procedures run to method specific fixed points.

A third strand addresses deployment. National and sectoral systems such as the Romanian Netting service demonstrate that cycle-based compensation can be implemented at scale, but they also highlight the coordination burden associated with cycle discovery and centralised visibility. More recent distributed ledger and consortium architectures improve traceability and access control, yet many published accounts remain architectural and provide limited evidence on network level obligation reduction. Related work on invoice finance blockchain systems often focuses on authenticity, anti-fraud controls, or receivables registration rather than on multilateral trade credit clearing itself.

Clearing also sits next to supply chain finance mechanisms that address late payments through liquidity rather than the reduction of gross obligations. Invoice discounting and factoring accelerate cash flow by bringing in financing before invoices mature. This adjacent line matters for clearing because it highlights the alternative lever used when netting is unavailable or insufficient: pay earlier by injecting liquidity, rather than cancel obligations by matching them (Fabrizio et al., 2019; Wehinger, 2014).

Across research and practice, scalable automation for clearing and netting has followed three broad directions.

First, centralised optimisation treats debt clearing as a constrained optimisation problem. Exact methods such as dynamic programming can provide optimal solutions for restricted instances, but their computational growth limits them to smaller graphs (Pătcaş, 2011). Heuristic approaches, including evolutionary search, improve runtimes but do not guarantee optimality and still assume a single solver with visibility into the full obligation graph (Pătcaş and Bartha, 2019).

Second, deployment driven national scale systems use graph routines that trade optimality for speed and operational reliability. The Romanian Ministry of Economy Netting service is among the best documented nation scale implementations (Gavrilă and Popa, 2021). In the reported workflow, the system partitions payables into strongly connected components, cancels debt on directed cycles, and iterates until no positive weight circuit remains. A key limitation is architectural: central operation requires firms to disclose invoice information into a single pool, which can become a coordination bottleneck and concentrates operational risk.

Third, decentralised or consortium-based protocols replace the single hub with permissioned shared execution. Examples include CLSNet for foreign exchange payment netting and Hyperledger based proofs of concept for inter carrier settlement (CLS Group, 2018; Hyperledger Telecom Special Interest Group, 2019). These systems demonstrate auditability and access control, but public reporting often provides limited network level evidence on the reduction on gross obligations ratios and realised debt relief in typical trade credit graphs.

Beyond netting workflows, blockchain research in invoice finance has highlighted another cluster of objectives: preventing invoice fraud and reducing verification costs in discounting and factoring. Factoring systems must handle authenticated identities and risks such as double pledging, and confidentiality pressures can block the emergence of a single shared invoice registry (Battaiola et al., 2019). Several designs propose smart contract-based registration of factoring agreements to prevent double factoring while preserving auditability and dispute handling (Mohammadzadeh et al., 2021). In discounting settings, blockchain based approaches also emphasise time stamped records of invoice existence, debtor receipt, and verification, aiming to reduce audits and reconciliation effort (Fabrizio et al., 2019). A more direct connection to trade credit clearing is provided by Bottazzi et al. (2024), who study multilateral trade credit set-off under privacy constraints using secure computation and graph anonymization. Their formulation is especially relevant because it treats trade credit set-off itself, rather than invoice registration or generic netting architecture, as the primary object of analysis. In that sense, it is closer to the present paper than blockchain invoice finance registries, while still differing from our approach in method and deployability assumptions: their focus is privacy preserving optimisation with stronger computation and coordination requirements, whereas Byppay is designed as a local greedy reduction rule that can be executed with limited neighbourhood disclosure.

Recently, open decentralised clearing protocols have been proposed to extend clearance beyond closed clubs while preserving privacy. The Cycles Protocol positions itself as an open decentralised clearing, settlement, and issuance system that aims to clear more debt with less money through a privacy preserving multilateral settlement platform based on graph optimisation (Buchman et al., 2024). Its motivation is explicit about why small firms stay out of clearing clubs: exposing financial relationships without strong rules and privacy protections can be risky. Cycles Protocol (Buchman, et al., 2024) proposes atomic multilateral operations executed via distributed consensus and sketches a privacy approach that combines trusted execution environments for computation with zero knowledge proofs for integrity, so that the system can verify balanced settlement outcomes even under adversarial conditions. At the algorithmic level, it emphasises extracting liquidity from cyclic structures in the payment network, which aligns with the broader cycle-oriented settlement intuition.

And, to ground the discussion in operational scale, the Romanian circuit based Netting system has reported cumulative processing at country level over multiple years. Table 1 summarises the reported volumes for the period covered by the service documentation.

**Table 1** – Historical outputs of the Romanian Ministry of Economy Netting service under manual operation, from 2000 to 2017.

| Year | Number of circuits | Amounts (RON) | Average exchange rate EUR-RON | Amounts (billion EUR) | GDP (billion EUR) | % of GDP |
|---|---|---|---|---|---|---|
| 2000 | 5,585 | 2,403,053,727 | 1.9955 | 1.2 | 40.3 | 2.99 |
| 2001 | 42,652 | 12,421,327,031 | 2.6026 | 4.8 | 44.9 | 10.63 |
| 2002 | 47,034 | 19,911,421,429 | 3.1255 | 6.4 | 48.5 | 13.14 |
| 2003 | 52,436 | 22,112,743,029 | 3.7555 | 5.9 | 52.6 | 11.19 |
| 2004 | 49,863 | 22,801,524,621 | 4.0532 | 5.6 | 60.8 | 9.25 |
| 2005 | 33,974 | 17,779,815,045 | 3.6234 | 4.9 | 79.5 | 6.17 |
| 2006 | 24,535 | 14,848,420,479 | 3.5245 | 4.2 | 97.7 | 4.31 |
| 2007 | 16,250 | 12,041,806,899 | 3.3373 | 3.6 | 123.7 | 2.92 |
| 2008 | 14,215 | 13,869,438,123 | 3.6827 | 3.8 | 139.7 | 2.70 |
| 2009 | 17,066 | 15,876,671,658 | 4.2373 | 3.7 | 118.3 | 3.17 |
| 2010 | 18,432 | 15,159,786,445 | 4.2099 | 3.6 | 124.1 | 2.90 |
| 2011 | 18,311 | 15,715,962,869 | 4.2379 | 3.7 | 131.5 | 2.82 |
| 2012 | 15,018 | 15,246,545,476 | 4.4560 | 3.4 | 133.9 | 2.56 |
| 2013 | 10,643 | 12,774,006,762 | 4.4190 | 2.9 | 144.7 | 2.00 |
| 2014 | 8,400 | 9,929,328,661 | 4.4446 | 2.2 | 150.8 | 1.48 |
| 2015 | 7,117 | 8,371,338,603 | 4.4450 | 1.9 | 159 | 1.18 |
| 2016 | 5,719 | 6,377,781,136 | 4.4908 | 1.4 | 185.7 | 0.76 |
| 2017 | 4,218 | 5,720,174,633 | 4.5682 | 1.3 | 182.1 | 0.74 |
| **Total** | **391,468** | **243,361,146,626** | | **64.6** | | |

Table 1 summarises the reported annual volumes processed by the Romanian Ministry of Economy Netting service during its manual operation period, from 2000 to 2017. Amounts are reported in RON and converted to euros using the annual average EUR RON exchange rate shown. Over this period, the service processed 391,468 circuits and a cumulative 64.6 billion euros equivalent of invoice amounts. This historical record provides context for the scale and relevance of cycle-based netting in a national setting.

Gavrilă and Popa (2021) describe an enhanced implementation of the same netting intuition that was algorithmically deployed from 2020 onward and benchmarked with the manual implementation of the said cycle-based compensations. In this paper, we use the Netting routine described in that work as the next cycle-based baseline. Regarding the 2021 invoice corpus of 133,191 invoices totalling €19,669,036,729.80 that is obtained from the IMI service webpage https://gama.cppi.ro . On this corpus, Netting reports the cancellation of €4,128,146,348.60, which corresponds to a 20.99% reduction in total outstanding invoice obligations under the evaluation metric defined in Section 4.

The annual 2021 debt graph behind that data is structurally sparse but economically connected (see Table 2). Aggregating the 133,191 invoice records into firm to firm directed obligations yields a weighted network of 11,725 firms linked by 14,739 unique directed relations, with density of only 0.0107% and sparsity of 99.9893%. This confirms that the empirical setting is highly sparse in the graph theoretic sense, which is consistent with Gavrilă and Popa's characterization of invoice networks as sparse structures in which firms transact with only a limited subset of potential counterparties. Despite this sparsity, the network is not economically fragmented. The largest weakly connected component contains 10,685 firms, or 91.13% of all firms, while the largest strongly connected component contains 1,534 firms, or 13.08% of the annual graph, and 22.58% of all firms belong to nontrivial strongly connected components. Reciprocity is also nonnegligible, with 32.62% of directed links reciprocated. These features imply that the 2021 archival market combines a thin overall graph with a large connected trading backbone, and a meaningful reciprocal core in which circular compensation can occur. The weighted structure is even more concentrated. The median directed edge is only €2,231.54, whereas the mean edge weight is €1.334 million and the edge weight Gini coefficient is 0.9687, indicating a highly unequal obligation structure dominated by a relatively small set of very large links. In substantive terms, the 2021 network is therefore best understood as a sparse but connected and strongly concentrated debt graph, with enough reciprocal structure to support compensability mechanisms while remaining far from a dense or homogeneous market.

**Table 2**. Structural characteristics of the 2021 archival debt graph

| **Measure** | **Notation** | **Value** |
|---|---|---|
| Number of firms | $N$ | **11,725** |
| Number of directed links | $M$ | **14,739** |
| Annual invoice count | | **133,191** |
| Annual invoice volume | | **€19,669,036,729.80** |
| Directed density | $M/[N(N-1)]$ | **0.0001072** |
| Directed density (%) | | **0.0107%** |
| Sparsity | $1 - M/[N(N-1)]$ | **0.9998928** |
| Sparsity (%) | | **99.9893%** |
| Reciprocity | reciprocated directed links / all directed links | **0.3262** |

| Measure | Notation | Value |
|---|---|---|
| Reciprocity (%) | | **32.62%** |
| Mutual dyads | | **2,404** |
| Average in degree | $k^{in}$ | **1.2571** |
| Average out degree | $k^{out}$ | **1.2571** |
| Average total degree | $\bar{k}$ | **2.5141** |
| Median in degree | | **1** |
| Median out degree | | **0** |
| Maximum in degree | | **453** |
| Maximum out degree | | **6,050** |
| Weakly connected components | WCC | **325** |
| Largest weakly connected component | largest WCC | **10,685** |
| Largest WCC share | largest WCC / N | **91.13%** |
| Strongly connected components | SCC | **9,394** |
| Largest strongly connected component | largest SCC | **1,534** |
| Largest SCC share | largest SCC / N | **13.08%** |
| Nodes in nontrivial SCCs | SCC size >1 | **2,647** |
| Share in nontrivial SCCs | | **22.58%** |
| Undirected transitivity | | **0.0000715** |
| Undirected average clustering | | **0.0161** |
| Number of undirected triangles | | **468** |
| Maximum k core | | **5** |
| Size of maximum k core | | **49** |
| Degree assortativity | | **-0.3870** |
| Mean edge weight | | **€1.334m** |
| Median edge weight | | **€2,231.54** |
| 90th percentile edge weight | | **€736,017.99** |
| 99th percentile edge weight | | **€17.094m** |
| Gini coefficient of edge weights | | **0.9687** |
| Mean invoices per directed link | | **9.04** |
| Median invoices per directed link | | **1** |
| Maximum invoices on one directed link | | **5,626** |
| Gini coefficient of out strength | | **0.9831** |
| Gini coefficient of in strength | | **0.9873** |

**Note**: The graph is constructed by aggregating the 2021 invoice level ERP records into a directed weighted annual network. Firms are nodes, and each directed edge represents the total invoiced obligation from one firm to another over the year. The 2021 setting is modelled as the archival invoice network of 11,725 firms matched to realised Netting and Byppay outcomes, consistent with the invoice graph representation described by Gavrilă and Popa in the Romanian clearing environment.

The potential improvement of local path compensation is structural: Cycle-based netting can only cancel obligations that can be reduced through directed cycles available to the cycle finding procedure. The approach we propose here, Byppay, applies a local three-party step A → B → C (referred to as *local path step* or simply *Byppay step*) that can also cancel along open paths by pushing a common amount through an intermediary. Since open chains are common in supply chain invoice networks, local path steps can reduce obligations even when cycle opportunities are limited.

It is also useful to separate auditability from structural clearing capacity. Moving a clearing workflow onto a blockchain can improve auditability, integrity, and coordination, but it does not change the settlement primitive. For Netting, a ledger can record cycle settlement steps, but the method still relies on the existence of directed cycles and on some form of coordination to select and execute cycles. For Cycles Protocol, the ledger is central to coordinating private intent posting and verifying batch outcomes, but the clearing intuition remains circuit oriented and the protocol introduces batch execution constraints. For local path compensation, a blockchain can act as an auditable bulletin board for payment instructions and confirmations. Because the computation is local, it does not require pooled disclosure of the full obligation graph and it can begin with smaller adoption clusters where local paths already exist.

A similar separation appears in invoice finance. Blockchain can strengthen proof of existence, timestamping, and verification for invoices, and it can reduce manual checks in discounting workflows, but it does not by itself change whether a clearing algorithm is limited to cycles or can propagate along paths (Fabrizio et al., 2019).

# 3. Models and Definitions

We are comparing two families of settlement logic applied to the same invoice debt graph. Both start from identical data: a directed weighted graph where each node is a company and each directed edge represents the amount a debtor owes a creditor, aggregated from invoices for a given month.

The first family is **circuit-based compensation**, represented by the Netting approach. It searches for directed cycles in the debt graph and settles them by offsetting obligations around the loop. The method is fundamentally cycle limited: only debts that lie on at least one directed circuit can be reduced. In most implementations it also relies on global visibility to enumerate circuits and coordinate which ones are executed.

The second family is **local path compensation**, represented by the algorithm of this paper, Byppay. As introduced, instead of looking for cycles, it performs a local three-party reduction A→B→C whenever *A* owes *B* and *B* owes *C*. It transfers the matched amount as a direct payment instruction from A → C and decreases both adjacent debts accordingly. Because it does not require cycles, it can reduce obligations along open chains while preserving node net positions on the combined post-settlement state. This local structure also makes it *agentifiable*: each firm can compute and execute steps based on its own receivables and payables plus limited bilateral coordination, as a ready protocol for decentralised platforms.

The two families are formalised below

## 3.1. Problem Setting and Notation

We model an invoice set as a directed weighted debt graph $G = (V, E, d)$. Consider one evaluation window $m$, typically one calendar month, and denote the corresponding graph by

$$G_m = (V, E_m, d_m)$$

$V$ is the set of companies. $E \subseteq V \times V$ are the directed obligation edges. A directed edge $(u, v) \in E_m$ exists when company $u$ owes company $v$ , and weight $d\,(u, v) > 0$ is the aggregate unpaid amount, after invoices aggregation.

For each node $v$, define total incoming and outgoing obligations by

$$I(v) = \sum_{u \in V} d(u,v) \ , \qquad O(v) = \sum_{w \in V} d(v,w)$$

And net position

$$b(v) = I(v) - O(v)$$

Thus $b(v) > 0$ means that $v$ is a net receiver, $b(v) < 0$ means it is a net payer

The total invoice mass of a graph $G$ is:

$$D(G) = \sum_{(u,v) \in E} d(u,v)$$

We also track the number of positive obligation edges $|E|$ as a simple structural measure of graph simplification.

A settlement procedure takes $G$ as input and returns a **post settlement state**

$$S' = (G', \mathcal{P}),$$

where $G'$ is the residual invoice graph and $\mathcal{P}$ is the multiset of generated settlement promises to pay. Under this interpretation, Byppay does not by itself novate the original invoice claims. Rather, the invoice graph records residual original obligations, while $\mathcal{P}$ records the settlement instructions generated by path compression. Net position preservation is therefore defined on the combined state $S'$, not on the residual invoice graph alone.

For every node $v$, let $P_{in}(v)$ and $P_{out}(v)$ denote the total value of generated settlement promises entering and leaving $v$. The post settlement net position is

$$b'(v) = IG'(v) + P_{in}(v) - OG'(v) - P_{out}(v).$$

A valid settlement procedure preserves node net positions if

$$b'(v) = b(v) \text{ for all } v \in V.$$

The economically relevant residual quantity is not the residual invoice mass alone, but the **post settlement payable mass**

$$\Omega(S') = D(G') + P_{tot} \ ,$$

where

$$P_{tot} = \sum_{p \in P} value(p).$$

This is the amount that still needs to be discharged after clearing, whether through residual invoice obligations or through generated settlement promises.

Accordingly, for each evaluation window $m$, we define both **gross obligation reduction** and **liquidity relief** on the same economic basis:

$$GOR_m = LR_m = D(G_m) - \Omega(S'_m).$$

Hence, in this revised formulation, gross obligation reduction and liquidity relief are identical by construction, because both measure the reduction in economically relevant payable mass.

For Netting, no new settlement promises are generated, so $P_{tot,m}=0$, and therefore

$$GOR^N{}_m = LR^N{}_m = D(G_m) - D(G'^N{}_m).$$

For Byppay,

$$GOR^{B}{}_{m} = LR^{B}{}_{m} = D(G_m) - [D(G'^{B}{}_{m}) + P_{tot,\, m}].$$

We separately define **invoice layer compression** as

$$IC_m = D(G_m) - D(G'_m).$$

This measures how much weight disappears from the original invoice graph itself. For Byppay, $IC_m$ can exceed debt relief because one compensated tranche may be rerouted into a direct settlement promise rather than fully extinguished as post settlement payable mass. For that reason, $IC_m$ is reported only as a structural graph simplification metric and is not used as the primary economic outcome.

## 3.2 Formal algorithm for Netting: Circuits/cycles

This family of methods reduces debt only when it lies on a directed circuit or cycle. A directed cycle is a sequence of distinct vertices $(v_1, v_2, \ldots, v_k, v_1)$ such that every edge $(v_i, v_{(i+1)})$ exists and has positive weight.

Given a cycle $\zeta$, the cycle settlement amount is the minimum edge weight on that cycle:

$$\delta = \min_{(u,v)\in\zeta} d(u,v)$$

Then every edge on the cycle is reduced by $\delta$:

$$d(u,v) \leftarrow d(u,v) - \delta \text{ for all } (u,v) \in \zeta$$

Any edge that reaches zero is removed from $E$. This is the standard way to cancel debt around a closed loop: everyone in the loop pays a bit less ($\delta$), and the minimum edge on the loop is fully eliminated.

At scale, the method requires global visibility to discover cycles and coordinate which cycles to execute. Gavrilă and Popa' (2021) circuit settlement optimization algorithm reflects this by searching over candidate circuits and excluding incompatible choices.

Stopping rule used in the benchmark is when $G$ has no directed cycle. The netting run continues until no directed cycle remains in the positive debt graph. Formally, it stops when the remaining graph is acyclic.

For cycle netting, gross obligation reduction = liquidity relief (GOR = LR)

An example of the algorithm's processing and treatment of circuits, as described by Gavrilă and Popa (2021), is as follows:

There are eight companies (A, B, C, D, E, F, G, and H) that have invoiced each other, and the netting algorithm identifies circuits to apply full compensation of invoices. See the following plot that represents them:

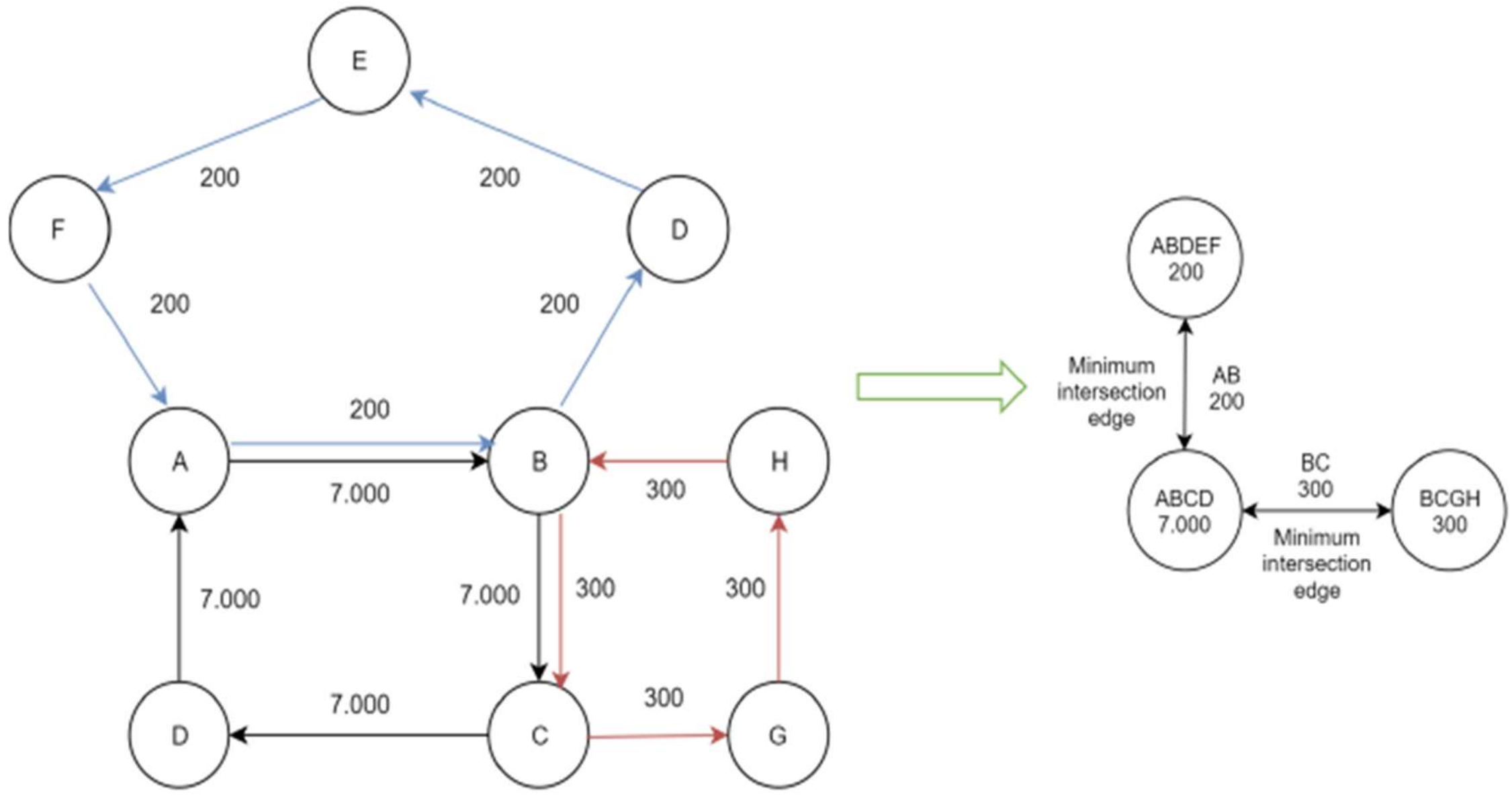

The netting algorithm identifies three circuits for full compensation of invoices. The first circuit consists of five companies (A, B, C, D, E, and F) and proposes a compensation of 200; the second circuit comprises four companies (A, B, C, and D) and proposes a compensation of 7,000; and the third circuit involves four companies (B, C, G, and H) and proposes a compensation of 300.

As a result of the netting process, the graph of debt shows a reduction in the number of arcs (from 11 to 2) and a decrease in global debt by 7,500.

```
Algorithm 1. Netting benchmark
Input:
   G0 = (V, E0, w0)     // original aggregated invoice graph
   V sorted in ascending node order
Output:
   GT                  // terminal residual invoice graph
   PT = 0              // empty settlement instruction layer
   L                   // ordered cycle log
Procedure:
   G <- copy of G0
   L <- empty list
   while G contains at least one simple directed cycle do
      gamma <- simple directed cycle with:
            1. minimum cycle length
            2. if tied, lexicographically smallest canonical sequence
      delta <- minimum residual edge weight on gamma
      for each directed edge e on gamma do
         w(e) <- w(e) - delta
         if w(e) = 0 then remove e from G
      end for
      append (gamma, delta) to L
   end while
   return (G, 0, L)
```

## 3.3 Formal algorithm for Byppay as Path Enabled (Local Three-Party Steps)

This algorithm, Byppay, does not require cycles. It performs local three-party compensation across an intermediary $B$ using any positive incoming and outgoing debts of $B$. A Byppay step is defined on a directed local 2-path A to B to C in the current residual invoice graph G.

For a node $B$, choose a client $A$ such that $d(A, B) > 0$ and a supplier $C$ such that $d(B, C) > 0$.

Define the local compensation amount:

$$d = min(d(A, B), d(B,C))$$

Update the two adjacent debts:

$$d(A, B) \leftarrow d(A, B) - d$$

$$d(B, C) \leftarrow d(B, C) - d$$

The residual invoice graph is updated by subtracting d from both source edges. Any edge whose residual weight reaches zero is removed from the residual edge set. If A is different from C, the settlement instruction layer is updated by adding d to payments. If A equals C, the move reduces to bilateral cancellation around a two-edge reciprocal relation and no endpoint instruction is stored.

Record a settlement instruction of value $d$ from $A$ to $C$ denoted for example by $p2pay(A,C) \mathrel{+}= d$.

In the present paper, this record is interpreted as a settlement instruction layer rather than as requiring novation of the underlying invoice claim. Accordingly, the original obligations A to B and B to C remain the legal accounting reference until discharge, while the settlement layer records how payment is to be executed or coordinated. This interpretation is sufficient for the benchmark and preserves comparability with cycle-based compensation.

This step depends only on $B$ and its direct neighbourhood, so it is agentifiable: each company can compute and propose steps locally with limited bilateral coordination with its direct counterparties.

Let $\mathcal{P}$ be the multiset of generated settlement promises to pay, and let

$$P_{tot} = sum(\mathcal{P}) = \sum_{p \in P} value(p)$$

If $G_{rem}$ denotes the terminal residual invoice graph, then the post settlement payable mass is

$$\Omega(S') = D(G_{rem}) + P_{tot}$$

The economically relevant outcome of a Byppay run is therefore

$$D(G) - \Omega(S')$$

which in this paper is reported both as gross obligation reduction and as liquidity relief.

The Byppay run continues until no further pairwise reductions are possible, meaning there is no node $B$ with at least one positive incoming edge that at least one positive outgoing edge can be matched.

A fixed-point condition is:

$$\text{For every } B \in V, \text{ either } In(B) = \emptyset \text{ or } Out(B) = \emptyset$$

where $In(B) = \{A : d(A,B) > 0\}$ and $Out(B) = \{C : d(B,C) > 0\}$.

Each Byppay step preserves combined net positions and removes at least one residual invoice edge, because one of the two reduced edges reaches zero by construction. The procedure therefore terminates after at most $|E|$ executed steps. If a node has at least one positive incoming and one positive outgoing edge, then a Byppay step exists at that node. Thus, termination means there is no such node left. A useful interpretation is that the remaining graph is directionally separated: each company is only a net payer or only a net receiver in what remains, so no local pass through compensation is possible.

**Remark. Settlement interpretation and relation to cycle compensation:** Bilateral netting is the two-edge special case of the local path rule. More generally, directed cycle compensation can be represented within the same settlement logic as a special case in which the path closes back to its starting firm. Under a settlement instruction or promise to pay interpretation, the essential difference between Netting and Byppay is therefore not that one method generates settlement records while the other does not, but that their admissible move sets differ. Netting is restricted to obligations embedded in directed cycles, whereas Byppay can also act on open paths.

This "until no more moves exist", run-to-fixed-point design makes the two procedures comparable, because both are evaluated as greedy reduction processes under their respective admissible move sets. The existence of a fixed point does not by itself imply uniqueness: different execution orders can lead to different residual graphs and different sets of settlement instructions, even when each node's net position is preserved. To ensure reproducibility, we therefore run each procedure with a deterministic tie breaking policy defined below.

Deterministic execution policy for Byppay is the following: we iterate intermediaries $B$ in a fixed order (e.g., increasing node identifier). For each $B$, we select $A$ from incoming neighbours and $C$ from outgoing neighbours using a deterministic rule (e.g., highest weight first, with ties broken by node identifier), then apply $d = min(d(A,B), d(B,C))$. We repeat until no eligible triple remains. For circuit netting, we likewise use a deterministic rule to select the next cycle (e.g., first cycle found by a fixed traversal order) and apply the minimum edge cancellation. This makes the benchmark results repeatable and comparable across runs.

The economically relevant outcome of a Byppay run is the reduction in post settlement payable mass, that is, D(G)−D(G(S′), which in this paper is reported both as gross obligation reduction and as liquidity relief. The underlying local debt simplification primitive is known but not previously formalized and benchmarked in the invoice network setting, to our knowledge. The contribution of this paper is then a novel usage of an algorithm with its adaptation to invoice backed trade networks, its reproducible benchmarking against a real cycle-based baseline, and its deployment framing for decentralisable clearing under selective disclosure constraints, that is novel. It performs a local three party step across an intermediary *B*. Choose *A* and *C* such that given debt $d_{AB} = d(A,B) > 0$ and $d_{BC} = d(B,C) > 0$, and reduce both debts by $d=min(d(A,B), d(B,C))$, and record a payment instruction from *A* to *C*. The run continues until no node has both a positive incoming and a positive outgoing edge that can be matched, so the remaining graph becomes directionally separated.

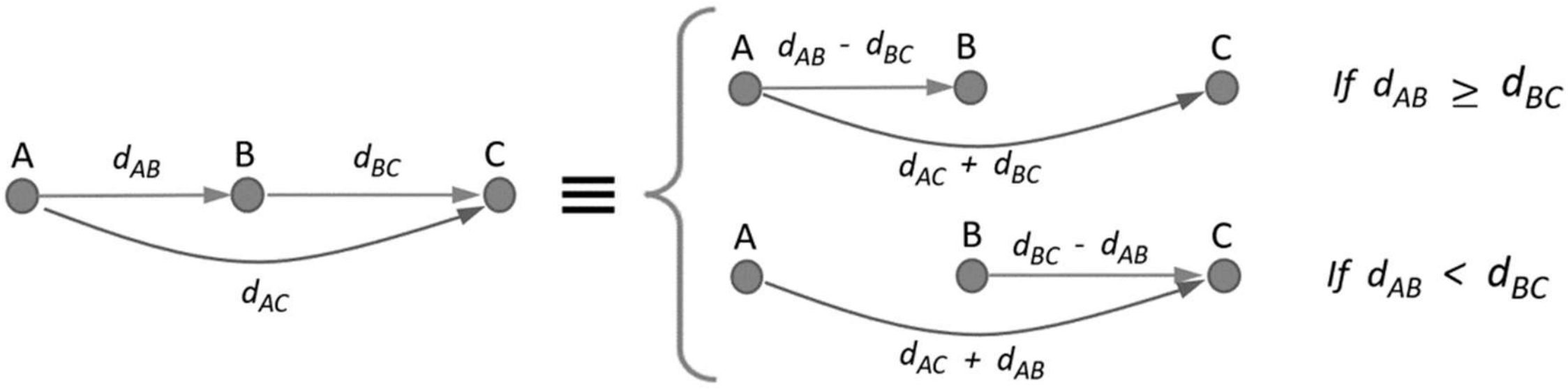

**Fig 1.** Graph representation of 3-edge before and after a local path step compensation, *Byppay step.*

Byppay is not designed to compute a globally optimal clearing plan. It applies a local three-party reduction and terminates at a fixed point defined by the absence of any further local matches. In the aggregated invoice graph, each executed step reduces two adjacent edges by *s* and sets at least one of them to zero, so the number of steps is at most the number of positive edges in the initial graph. With local adjacency lists, a valid step can be proposed and verified using only neighbourhood information, so the procedure is efficient and compatible with decentralised execution.

The idea of path or chain contraction or shortening is algorithmically known but novel in this application. The value here is mostly in packaging it as a decentralisable rule set and testing it on a large real invoice corpus against one specific operational baseline.

This local move set has a direct deployment implication. Each step depends only on *B* and its direct neighbourhood, so each company can compute and propose steps locally with limited bilateral coordination with direct counterparties. It is a local path compensation.

In other words, this new approach can deliver value without building a pooled global obligation graph, because it is a local path algorithm for invoice backed debt graphs. Byppay treats intermediaries such as *B* in a path A→B→C as pass through nodes. By converting part of an indirect chain into a direct settlement promise from *A* to *C*, the algorithm can propagate reductions through the network until the graph reaches a state of directional separation. In that terminal state, each remaining node is only a net payer or only a net receiver in the residual graph, so no further local path reduction is possible.

A 2-edge cycle is the special case in which the Byppay step reduces to bilateral netting, the same way the cycle restricted Netting does.

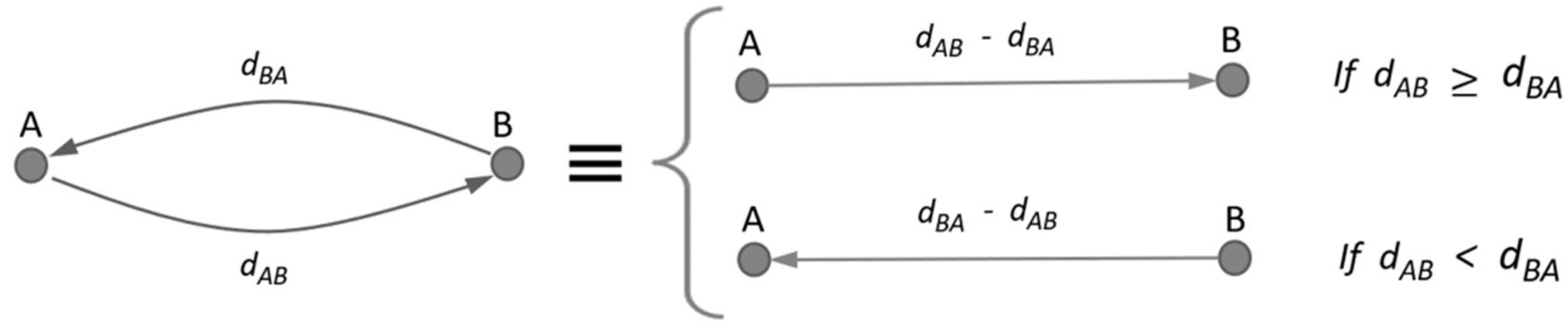

**Fig 2.** Graph representation of 2-edge before and after a Byppay step

```
Algorithm 2. Byppay
Input:
   G0 = (V, E0, w0)     // original aggregated invoice graph
   V sorted in ascending node order
Output:
   GT                  // terminal residual invoice graph
   PT                  // terminal settlement instruction layer
   L                   // ordered step log
Procedure:
   G <- copy of G0
   P(u, v) <- 0 for all (u, v) in V x V
   L <- empty list
   while exists b in V such that InG(b) and OutG(b) are both nonempty do
      b <- smallest eligible intermediary
      a <- incoming neighbor of b with maximum residual weight
          ties broken by smaller node identifier
      c <- outgoing neighbor of b with maximum residual weight
          ties broken by smaller node identifier
      d <- min(w(a, b), w(b, c))
      w(a, b) <- w(a, b) - d; if zero remove (a, b)
      w(b, c) <- w(b, c) - d; if zero remove (b, c)
      if a != c then P(a, c) <- P(a, c) + d
      append (a, b, c, d) to L
   end while
   return (G, P, L)
```

**Example: one Byppay step on a three-firm path**

Consider three firms (A), (B), and (C), with two invoice edges:

*A to B: x,* *B to C: y.*

Let

$$d = min(x, y).$$

One Byppay step applies on the path A to B to C. It reduces both invoice edges by *d* and records a settlement promise from A to C of value *d*. After the step,

*A to B : x − d,* *B to C: y − d,*

and

*p2pay(A,C) += d.*

The initial invoice mass is

$$D(G) = x + y$$

*The residual invoice mass is*

$$D(G') = (x - d) + (y - d) = x + y - 2d$$

The total value of generated settlement promises is

$$P_{tot} = d$$

Hence the post settlement payable mass is

$$m = D(G') + P_{tot} = (x - d) + (y - d) + d = x + y - d$$

Therefore, the economically relevant relief is

$$GOR = LR = D(G) - \Omega(S') = (x + y) - (x + y - d) = d$$

By contrast, the larger quantity

$$IC = D(G) - D(G') = 2d$$

is **not** debt relief. It is only invoice layer compression, meaning the amount of weight removed from the original invoice graph.

For example, if x = y = 100, then initially D(G) = 200. After one Byppay step, the residual invoice graph has zero mass, but a settlement promise A to C of 100 remains. Thus

$$\Omega(S') = 100,$$

and therefore

$$GOR = LR = 200 - 100 = 100.$$

The quantity 200 is merely the compression achieved on the invoice layer. The liquidity relief is 100, because 100 of payable mass still remains in the form of the generated settlement promise. This distinction resolves the apparent contradiction in earlier formulations. Byppay can produce greater compression of the invoice graph than immediate liquidity relief, but once gross obligation reduction is defined on the combined post settlement state, it is exactly equal to liquidity relief.

## 4. The temporal and topological dimensions

Because invoices have due dates and new obligations continuously enter the debt graph, the comparison in this section incorporates a temporal dimension. Thus, this is a generalisation of section 3.1 where we were considering monthly graph update.

Remembering that $G = (V, E, d)$ is a directed invoice graph, where each node is a company and each directed edge $u \rightarrow v$ with weight $d(u, v) > 0$ means that $u$ owes $v$. A local path of length two is a triple A → B → C with $d(A, B) > 0$ and $d(B, C) > 0$. A directed circuit is a closed directed walk on distinct vertices. Byppay acts on local 2-paths through an intermediary $B$. Netting acts only on directed circuits.

For a dynamic invoice graph $G(t)$, let $L_B(t)$ and $L_N(t)$ denote the cumulative liquidity relief delivered by local path and by cycle up to time $t$. Throughout, due dates are treated as economically meaningful deadlines. Therefore, the relevant comparison is not only final cumulative relief but relief available before maturity.

The number of directed local 2-paths in the graph is:

$$P_2(G(t)) = \sum_{v \in V} d^-(v) \cdot d^+(v)$$

where $d^-(v)$ and $d^+(v)$ denote the in-degree and out-degree of node $v$, respectively. Each term counts the number of ordered pairs $(A, C)$ such that $A \to v \to C$, that is, the number of local 2-paths passing through intermediary $v$.

Define

$P_2°(G(t)) = \{ (A,B,C) \in V^3 : (A,B) \in E(t), (B,C) \in E(t),$ and the 2-Path $A \to B \to C$ lies on at least one directed cycle$\}$

Equivalently, $P_2°(G(t))$ counts those local 2-paths whose two edges belong to the same strongly connected region of the graph, so that there exists a directed return path from $C$ back to $A$. This is the local support of cycle restricted netting.

Therefore,

$P_2°(G(t)) \leq P_2(G(t)),$

with strict inequality whenever the graph contains at least one open local 2-path that does not lie on any directed cycle. The quantity $P_2(G(t))$ counts all directed local 2-paths and $P_2°(G(t))$ as the number of local 2-paths that lie on at least one directed circuit.

## 4.1. Temporal prevalence under due dated invoices

Let's formulate the following:

**Theorem 1. *Temporal prevalence of Byppay before cycle closure***

Consider a dynamic invoice graph $G(t)$. Suppose that at time $T_1$ there exists a directed path A → B → C with positive debts $d_{AB}(T_1) > 0$ and $d_{BC}(T_1) > 0$, but there is no closing edge C → A at $T_1$. Let $d_1 = min\{d_{AB}(T_1), d_{BC}(T_1)\}$ and assume that the closing edge C → A appears only later, at time $T_2 > T_1$. Then: (i) a Byppay step is feasible immediately at $T_1$ and yields liquidity relief $d_1$ ; (ii) any cycle restricted netting procedure yields zero relief on this subgraph for every $t$ in $[T_1, T_2)$; and therefore (iii) for every invoice due date $\tau$ in $[T_1, T_2)$, cumulative pre-maturity liquidity satisfies $L_B(\tau) - L_N(\tau) \geq d_1 > 0$.

***Proof.*** At time $T_1$, node $B$ has one positive incoming edge and one positive outgoing edge. By definition of the Byppay move, a local reduction is admissible with amount $d_1 = min\{d_{AB}(T_1), d_{BC}(T_1)\}$. This creates immediate liquidity relief $d_1$. By contrast, cycle restricted netting requires a directed circuit. Since the closing edge C → A is absent on the entire interval $[T_1, T_2)$, the subgraph remains acyclic there, so netting cannot execute any move on it. Hence $L_N(\tau) = 0$ and $L_B(\tau) \geq d_1$ for every $\tau$ in $[T_1, T_2)$, which gives $L_B(\tau) - L_N(\tau) \geq d_1 > 0$ □

**Corollary 1. *Economic significance of temporal prevalence***

If at least one invoice in the path A → B → C matures at some date $\tau$ in $[T_1, T_2)$, then Byppay can reduce payable pressure before maturity whereas cycle restricted netting cannot. Consequently, earlier path-based relief has strictly greater operational value, because it can prevent late payment, penalty costs, and downstream liquidity propagation that later cycle settlement cannot undo.

***Proof.*** Theorem 1 gives $L_B(\tau) - L_N(\tau) \geq d_1 > 0$ for every maturity date $\tau$ in $[T_1, T_2)$. Therefore, only Byppay can make liquidity available before the relevant due date. Once maturity is missed, later relief no longer has the same operational value because the late-payment event has already occurred □

**Corollary 2.** ***Integrated temporal advantage***

Over the interval $[T_1, T_2)$, the integrated temporal advantage of Byppay satisfies
$\int_{[T_1,T_2)} \left(L_B(t) - L_N(t)\right)dt \geq d_1 (T_2 - T_1)$.

***Proof.*** Theorem 1 gives $L_B(t) - L_N(t) \geq d_1$ for every $t$ in $[T_1, T_2)$. Integrating this pointwise lower bound over the interval yields the stated inequality □

## 4.2. Due date preservation

After stating temporal prevalence, it is useful to clarify the maturity of the settlement instruction generated by a local path step. In this paper, a Byppay instruction is interpreted as a settlement instruction or promise to pay attached to the matched compensated tranche, not as a new independent obligation with an arbitrary later maturity. Its due date should therefore inherit the temporal priority of the underlying matched obligations.

**Proposition. Due date preservation of the settlement instruction in a local path step.**
Consider a local path A → B → C in the invoice graph, with positive debts

$$d(A,B)>0\,,\quad d(B,C)>0,$$

and corresponding due dates

$$\tau_{AB}\;,\quad \tau_{BC}\,,$$

Let

$$d = min\{d_{AB},\, d_{BC}\}$$

A Byppay step reduces both debts by $d$ and records a settlement instruction, or promise to pay, of amount $d$ from A → C .
If the due date of that settlement instruction is defined by

$$\tau_{AC} := min\{\tau_{AB},\, \tau_{BC}\}$$

then the path-based settlement instruction preserves the temporal priority of the compensated tranche. In particular, it does not introduce a new economic maturity later than the one already governing the matched obligations.

Moreover, if $\tau_{AB} = \tau_{BC} = \tau$ , then the settlement instruction A → C also has due date $\tau$. Hence, in the equal due date case, the path-based settlement instruction has the same due date as the compensated amount would have in a directed circuit built on the same matched tranche.

Proof is straightforward.

Hence the settlement instruction inherits exactly the same due date as the two matched obligations. This is the same maturity logic that applies in a circuit when all compensated edges share a common due date: the compensated amount is settled relative to that existing due date rather than through the creation of a later one. Therefore, the path-based promise to pay preserves the temporal status of the compensated amount. □

**Corollary 3. Equal due date alignment between path and circuit settlement.**
Suppose a directed circuit contains a matched compensated tranche $d$ supported by obligations that all share the same due date $\tau$. Then the compensated amount in the circuit and the promise to pay generated by the corresponding local path interpretation both have due date τ. Therefore, under equal due dates, the temporal logic of path-based settlement is identical to that of circuit compensation.

**Proof.** Immediate from the proposition, since in the equal due date case the promise to pay inherits the same common maturity τ. □

The key temporal advantage of Byppay is not that it creates a later or artificial obligation, but that it can activate settlement on an already existing due dated economic flow before a directed cycle has formed.

## 4.3. Topological prevalence in invoice networks

Temporal prevalence states that Byppay can act earlier. Topological prevalence states that Byppay can act on a structurally larger opportunity set. This follows because the natural move set of Byppay is the set of all local directed n-paths, whereas the natural move set of Netting is only the subset of those n-paths that lie inside the cyclic core of the invoice graph. For notational simplicity, write $G$ for $G(t)$ in the discussion that follows.

$$q(G) = P_2^{\circ}(G) / P_2(G), \quad 0 \le q(G) \le 1$$

$$R_{top}(G) = P_2(G) / P_2^{\circ}(G) = 1 / q(G) \text{ whenever } P_2^{\circ}(G) > 0$$

The scalar $q(G)$ is the cycle-closure ratio, namely the fraction of local paths that are actually cyclic. Its reciprocal $R_{top}(G)$ is the topological opportunity ratio. Low $q(G)$ means a path-rich but cycle-poor invoice graph. High $q(G)$ means that most local paths already lie in the cyclic core.

**Theorem 2.** ***Topological prevalence of local paths over circuits***

For every directed invoice graph $G$: (i) Byppay is applicable exactly on the support counted by $P_2(G)$; (ii) cycle restricted netting is applicable only on the support counted by $P_2^{\circ}(G)$; and therefore (iii) $P_2^{\circ}(G) \le P_2(G)$, with strict inequality whenever $G$ contains at least one open local 2-path that does not belong to any directed circuit.

***Proof.*** Byppay requires a node with at least one positive incoming edge and at least one positive outgoing edge. That condition is equivalent to the existence of a directed local 2-path $A \rightarrow B \rightarrow C$. Hence the support of Byppay is counted exactly by $P_2(G)$. Netting requires a directed circuit, so only those local 2-paths that lie on at least one directed circuit can contribute to its support; these are counted by $P_2^{\circ}(G)$. Every cyclic n-path is a n-path, but an open n-path need not be cyclic. Thus $P_2^{\circ}(G) \le P_2(G)$, and the inequality is strict whenever at least one open n-path exists outside the cyclic core □

**Corollary 4.** ***Acyclic invoice graphs***

If $G$ is acyclic, then $P_2^{\circ}(G) = 0$. Hence cycle restricted netting is inapplicable, while Byppay remains applicable whenever $P_2(G) > 0$.

***Proof.*** An acyclic directed graph contains no directed circuit by definition, so no local n-path can belong to a circuit. Therefore $P_2^{\circ}(G) = 0$. Byppay still applies whenever some node has both positive in-degree and positive out-degree, which is equivalent to $P_2(G) > 0$ □

## 4.4. Topology-dependent propositions

**Proposition 1.** ***Chains, trees, and layered supply chains***

If $G$ is a directed acyclic graph, then $P_2°(G) = 0$. In particular, for a directed chain on $n$ nodes, $P_2(G) = n - 2$ and $P_2°(G) = 0$. For a layered feed-forward supply chain $L_1 \to L_2 \to \cdots \to L_h$ with edges only from one layer to the next, $P_2(G)$ equals the sum, over internal layers $i = 2, \ldots, h-1$, of $|L(i-1)| \cdot |L(i)| \cdot |L(i+1)|$, and $P_2°(G) = 0$.

***Proof.*** Any directed acyclic graph contains no directed circuit, so $P_2°(G) = 0$. In a chain, each internal node contributes exactly one local 2-path, giving $P_2(G) = n - 2$. In the layered case, each node in an internal layer contributes one choice from the previous layer and one choice to the next layer, so the total number of local 2-paths is the sum, over internal layers $i = 2, \ldots, h - 1$, of $|L(i-1)| \cdot |L(i)| \cdot |L(i+1)|$. □

**Proposition 2.** ***Intermediary hub effect***

Suppose a node $B$ has $p$ distinct incoming neighbours and $q$ distinct outgoing neighbours. Then $B$ alone contributes $p \cdot q$ local 2-paths to $P_2(G)$. If no directed return path exists from any outgoing neighbour of $B$ back to any incoming neighbour of $B$, then none of these $p \cdot q$ local paths belongs to $P_2°(G)$.

***Proof.*** Each incoming neighbour of $B$ can be paired with each outgoing neighbour of $B$, yielding $p \cdot q$ distinct triples A → B → C. If no return path exists from any such $C$ back to any such $A$, then no triple A → B → C can lie on a directed circuit. Hence these $p \cdot q$ paths contribute to $P_2(G)$ but not to $P_2°(G)$ □

**Proposition 3.** ***Core–periphery structure***

Assume the invoice graph decomposes into a strongly connected core $H$ and an acyclic periphery attached in a feed-forward manner, so that no directed path from the periphery returns to close a cycle through $H$. Then $P_2°(G) = P_2°(H)$, whereas $P_2(G) \geq P_2(H) + P_2(\text{periphery})$. Consequently, the opportunity ratio $R_{top}(G)$ increases with the size of the open periphery.

***Proof.*** Under the stated assumption, any directed circuit must lie entirely inside the strongly connected core $H$. Therefore, only local 2-paths in $H$ can contribute to $P_2°(G)$, giving $P_2°(G) = P_2°(H)$. The periphery still contains directed local 2-paths, so its contribution is added to $P_2(G)$ but not to $P_2°(G)$. As the open periphery grows, the denominator of $q(G)$ stays anchored in the core while the numerator of $R_{top}(G)$ rises □

**Proposition 4.** ***Dense strongly connected subnetworks***

If every directed local 2-path in a subgraph lies on at least one directed circuit, then $P_2°(G) = P_2(G)$ on that subgraph and $R_{top}(G) = 1$. This is the topology most favourable to cycle restricted netting.

***Proof.*** The claim follows immediately from the definition of $P_2°(G)$. If each local 2-path is already cyclic, then the support of Byppay and the support of Netting coincide on that subgraph. In that case topological exclusivity disappears, although temporal prevalence may still arise in dynamic settings if the relevant circuit closes only later □

**Table 3** - Summary of topology classes

| **Topology class** | $P_2(G)$ | $P_2°(G)$ | $R_{top}(G)$ | **Implication** |
|---|---|---|---|---|
| Directed chain | n − 2 | 0 | +∞ | Only Byppay can act |

| Topology class | $P_2(G)$ | $P_2°(G)$ | $R_{top}(G)$ | Implication |
|---|---|---|---|---|
| Layered DAG | sum over internal layers | 0 | $+\infty$ | Large path supply, no netting |
| Hub without return paths | $p{\cdot}q$ at hub B | 0 | $+\infty$ | Multiplicative local-path effect |
| Core–periphery | $P_2(core)+$ $P_2(periphery)+$ … | $P_2°(core)$ | $> 1$ | Byppay acts in core and periphery |
| Dense strongly connected club | High | Near $P_2(G)$ | Near 1 | Netting becomes competitive |

## 4.5. Opportunity ratio and operation-count decomposition

Topology determines the raw supply of admissible local configurations, but the exact number of executed operations also depends on weights and deterministic tie-breaking. Still, one can write an exact decomposition of realised operation counts as a utilization factor times a topological opportunity count.

**Proposition 5.** ***Exact operation-count decomposition***

Let $K_B(G)$ and $K_N(G)$ be the numbers of executed Byppay and Netting operations (steps) under a fixed deterministic policy. Then $K_B(G) \le P_2(G)$ and $K_N(G) \le P_2°(G)$. Define the utilization factors $\alpha_B(G) = K_BG) / P_2(G)$ whenever $P_2(G) > 0$ and $\alpha_N(G) = K_N(G) / P_2°(G)$ whenever $P_2°(G) > 0$. Then

$$K_B(G) = \alpha_B(G)\, P_2(G)\,, \quad K_N(G) = \alpha_N(G)\, P_2°(G)$$

$$K_B(G) / K_N(G) = (\alpha_B(G) / \alpha_N(G))\, R_{top}(G) \quad \text{whenever } P_2°(G) > 0 \text{ and } K_N(G) > 0$$

***Proof.*** Each Byppay step is executed on a local 2-path and removes at least one of the two supporting edges, so that exact 2-path disappears permanently and no new invoice edge is created. Therefore, every Byppay step consumes at least one member of the initial support counted by $P_2(G)$, which gives $K_B(G) \le P_2(G)$. Similarly, each Netting step is executed on a directed circuit and reduces at least one edge on that circuit to zero; hence it permanently destroys at least one cyclic local 2-path from the initial support, and because no new invoice edge is created, $K_N(G) \le P_2°(G)$. The displayed equalities then follow by the definitions of $\alpha_B(G)$ and $\alpha_N(G)$, and the ratio identity is obtained by division □

**Corollary 5.** ***Low closure ratio implies many more Byppay applications***

If $\alpha_B(G)$ and $\alpha_N(G)$ are of the same order of magnitude and $q(G)$ is small, then the realised operation ratio $K_B(G) / K_N(G)$ is large. Equivalently, sparse feedback closure implies that Byppay can typically be applied many more times than Netting.

***Proof.*** By Proposition 5, $K_B(G) / K_N(G) = (\alpha_B(G) / \alpha_N(G)) / q(G)$. Holding the utilization-factor ratio away from zero and infinity, the operation ratio grows as $q(G)$ falls □

### 4.6. Illustrative examples

Let's see a number of examples below as concrete instantiations of the theorems and propositions above which together they show the difference between acting earlier, acting more often, and acting on a strictly larger structural opportunity set.

**Example 1. Dynamic three-company path that becomes a cycle**

Suppose that at time $T_1$ the invoice graph contains A → B with $d_{AB} = 80$ and B → C with $d_{BC}$ = 50, but no closing edge C → A. Then Byppay applies immediately with $d_1 = \min\{80, 50\} = 50$. Its pre-maturity liquidity relief at $T_1$ is therefore 50, while Netting provides 0 because there is no circuit.

Now suppose that at time $T_2$ a new edge C → A with $d_{CA} = 40$ appears. Netting can then execute on the 3-cycle A → B → C → A with cancellation amount $\delta = \min\{80, 50, 40\} = 40$. Since this is a 3-edge cycle, its immediate gross-obligation reduction and liquidity relief are both $3\delta = 120$. Byppay, however, had already relieved 50 before $T_2$. After the first step, the residual graph at $T_2$ contains A → B with weight 30 and C → A with weight 40, so Byppay can apply once more on C → A → B with amount 30. Byppay therefore reaches cumulative liquidity relief 80 by $T_2$. The important point is not equality at $T_2$, which need not hold, but strict temporal prevalence on every due date in $[T_1, T_2)$. If one averages cumulative relief over the interval, Netting yields (0+120)/2=60, whereas Byppay yields (50+80)/2=65. The precise averaging convention is not central here. The key point is that Byppay delivers strictly positive relief earlier, before the cycle closes.

**Example 2. Directed chain on five firms**

Consider 1 → 2 → 3 → 4 → 5. The internal nodes 2, 3, and 4 each support one local 2-path, so $P_2(G) = 3$. Because the graph is acyclic, $P_2°(G) = 0$. Netting cannot operate at all, while Byppay can act successively along the chain until no intermediary remains with both a positive incoming and a positive outgoing edge.

**Example 3. Three-layer feed-forward supply chain**

Let $|L_1| = 4$, $|L_2| = 3$, and $|L_3| = 5$, with every admissible invoice edge directed only from $L_1$ to $L_2$ and from $L_2$ to $L_3$ .Then

$$P_2(G) = |L_1|\,|L_2|\,|L_3| = 4 \times 3 \times 5 = 60, \quad P_2°(G) = 0$$

This is a compact illustration of why trade-credit networks can be rich in local paths yet poor in circuits: a modest feed-forward structure can already generate many local 2-paths and no cycles at all.

**Example 4. Intermediary hub**

Suppose an intermediary $B$ has $p = 4$ incoming debtors and $q = 5$ outgoing creditors. Then $B$ contributes $p{\cdot}q = 20$ local 2-paths by itself. If no return path exists from any outgoing neighbour back to any incoming neighbour, then all 20 are open paths. The local-path opportunity set is therefore positive and sizable, whereas the local cyclic opportunity set is zero.

**Example 5. Core–periphery invoice graph**

Suppose the core of the network is a trading triangle among three large firms and the surrounding periphery is a feed-forward set of supplier–customer chains attached to the core. The core contributes both local paths and cycles, but the periphery contributes only local paths. In such a graph, $P_2°(G)$ is determined essentially by the core, while $P_2(G)$ is augmented by the periphery. The opportunity ratio $R_{top}(G)$ is therefore greater than one and rises with peripheral breadth.

**Example 6: why the due date of the promise to pay in a path matches the due date used in a circuit**

Consider three firms A, B, and C, and suppose the following invoice obligations exist:

- A to B: 100, due on **30 June 2021**
- B to C: 100, due on **30 June 2021**

This is an open path A to B to C. Byppay applies with d = min {100, 100} = 100

The local path step reduces both invoice edges by 100 and records a settlement instruction, or promise to pay, from A to C of value 100.

The key point is that this promise to pay does **not** require a new economic maturity. The compensated amount is exactly the amount that (B) would have received from (A) and used to pay (C) on **30 June 2021**. Therefore, the natural due date of the settlement instruction (A \to C) is also **30 June 2021**.

So after the Byppay step:

- residual A to B = 0
- residual B to C = 0
- settlement instruction A to C = 100, due on **30 June 2021**

This shows that the promise to pay in the path does not introduce a different temporal obligation. It simply preserves the same settlement date that already governed the matched economic flow through the intermediary B.

Now consider the closed circuit:

- A to B: 100, due on **30 June 2021**
- B to C: 100, due on **30 June 2021**
- C to A: 100, due on **30 June 2021**

In a circuit compensation, the cancellation amount is: $\delta = \min\{100, 100, 100\} = 100$
All three obligations are reduced by 100. The economic effect is that the same amount that would have circulated through the loop on **30 June 2021** is extinguished on that same date.

Hence, in both cases:

- in the **circuit**, the compensated amount is tied to the common due date of the loop
- in the **path**, the promise to pay inherits that same due date from the matched obligations it replaces in settlement terms

So the path based settlement instruction is not temporally artificial. It carries the same maturity logic as the compensated amount in a circuit.

If there were unequal dates, suppose instead:

- A to B: 100, due on **20 June 2021**
- B to C: 100, due on **30 June 2021**

Then the matched amount can still be 100, but the settlement instruction A to C should not be given a later date than the economic flow it substitutes. A conservative rule is:

$$DueDate(A\ to\ C) = \min\{DueDate(A\ to\ B),\ DueDate(B\ to\ C)\}$$

So here the promise to pay (A to C) would be due on **20 June 2021**.

This makes the temporal logic precise: the settlement instruction inherits the earliest binding maturity of the obligations whose settlement it replaces. The same principle applies in a circuit, where the compensated tranche is understood relative to the due dated obligations that support the loop.

## 5. Benchmark design

The benchmark is conducted independently for each month of the 2021 data set and, separately, on the annual aggregate. For each evaluation window *m*, the raw input is an invoice table containing at least month, invoice identifier, debtor identifier, creditor identifier, and outstanding amount. Amounts are converted to a common currency and aggregated by ordered firm pair within each window. Only strictly positive outstanding amounts are retained.

Preprocessing is identical for both methods. The pipeline retains only positive records in the target window, removes records with missing debtor or creditor identifiers, excludes self-obligations, aggregates all invoices by ordered pair within the window, removes ordered pairs with zero aggregate amount, and maps firm identifiers to one fixed ascending node order.

For each month *m*, we build the debt graph $G_m$ by aggregating invoices by ordered pair (debtor, creditor). The monthly baseline amount is the total invoice mass

$$D(G_m).$$

Two independent copies of the same initial graph $G_m$ are then created. One copy is processed by Netting and the other by Byppay. No procedure is allowed to start from the residual graph of the other. A complete run returns the terminal residual invoice graph, the terminal settlement instruction layer, a step log recording every selected cycle or local path, the number of executed steps, and the terminal performance metrics.

Both procedures are run to a fixed point defined by their respective move sets. For Netting, the stopping condition is that no directed cycle remains in the positive debt graph. For Byppay, the stopping condition is that no node has both a positive incoming edge and a positive outgoing edge that can still support a local path reduction. This design measures how much each deterministic move set can reduce when repeatedly applied until no further admissible move exists.

The primary outcome metric is **liquidity relief**, defined as the reduction in post settlement payable mass. Because this is also the economically relevant notion of gross obligation reduction, we set

$$GOR_m = LR_m = D(G_m) - \Omega(S'_m),$$

where $S'_m$ is the terminal post settlement state of the relevant procedure.

For Netting, no new settlement promises are generated, so $\Omega(S'^N_m) = D(G'^N_m)$, and therefore

$$GOR^N{}_m = LR^N{}_m = D(G_m) - D(G'^N{}_m).$$

For Byppay,

$$\Omega(S'^B{}_m) = D(G'^B{}_m) + P_{tot,\, m},$$

and therefore

$$GOR^B{}_m = LR^B{}_m = D(G_m) - [D(G^B{}_m) + P_{tot,\, m}].$$

Thus the two procedures are compared on the same economic basis, namely the residual amount that still needs to be discharged after settlement.

We additionally report two secondary structural metrics. The first is **invoice layer compression**

$$IC_m = D(G_m) - D(G'_m),$$

which captures the amount of weight removed from the original invoice graph itself. The second is **edge reduction**, which measures simplification of graph topology through elimination of positive obligation links. These secondary metrics are useful for understanding structural simplification, but they are not substitutes for the primary economic outcome.

We also track operation counts:

$K^N{}_m$ = number of executed Netting cycle settlements,
$K^B{}_m$ = number of executed Byppay path steps.

Improvement ratios are computed on the primary economic metric. Annual totals for 2021 are obtained by summing monthly baseline amounts and monthly liquidity relief values. To summarise month to month variability, we report the arithmetic mean and standard deviation of monthly improvement percentages.

The benchmark therefore compares the two procedures under identical monthly invoice graphs, deterministic execution rules, and method specific fixed-point stopping conditions, while using one common economic metric. This avoids conflating graph compression with payable reduction and yields a like for like comparison of actual settlement relief.

## 5.1 One-Month Comparison

In April 2021, the dataset contains 11,488 invoices and €1,420,442,107 in total debt. Netting achieves €241,467,732.07 in compensation, equal to a 17% reduction, while Byppay achieves €881,397,558.00, equal to a reduction of up to 62.05%, or 3.65 times more. The main reason is structural. Netting only works when it can identify and execute directed settlement circuits, so debt embedded in open chains or tree like supplier structures cannot be compensated. Byppay does not require circuits, although a cycle can be interpreted as a special case of the broader local settlement logic when $C = A$. Instead, it performs local path step compensation and propagates reductions along paths, which allows it to capture compensation opportunities even in graphs where cycles are sparse. This also explains the stronger network simplification: Byppay executes 2,014 steps versus 642 circuits of Netting and achieves more graph density reduction, 25.29% versus 15.22%, meaning that more positive debt links are eliminated after settlement.

In a cycle-based settlement, existing obligations are reduced without changing the identity of the counterparties, while Byppay changes the obligation structure by replacing part of an indirect chain with a direct payment instruction/promise. Because each Byppay step depends only on a firm and its adjacent obligations, firms can execute it locally and concurrently with limited bilateral information exchange. By contrast, circuit-based approaches typically require global visibility to identify, enumerate, and coordinate cycles, which pushes the workflow toward centralised invoice collection and creates a coordination bottleneck.

## 5.2 A Full Year Comparison

As shown in Table 4, the 2021 dataset contains 133,191 invoices with a total invoiced amount of €19,669,036,729.80. The Netting algorithm compensates €4,128,146,348.60, equivalent to 20.99% debt relief. Byppay achieves €10,595,495,669 of liquidity relief, corresponding to 53.87% of the initial invoice mass. Under the revised metric convention adopted in this paper, this is also its gross obligation reduction. This means Byppay achieves 2.57 times the Netting result, or up to 156.66% more compensated debt.

Byppay continues to capture compensation opportunities even when cycles are sparse, which is typical of supply chain invoice graphs. This advantage is consistent across months. Byppay outperforms Netting every month, with monthly improvement ratios ranging from 149.10% in August to 441.22% in October. Although the monthly gains vary, the pattern is robust: the standard deviation of the monthly improvement ratios is 88.14%, and the annual aggregate remains clearly in Byppay's favour because the stronger reductions persist throughout the year.

In Table 4, the reported percentages for both algorithms are liquidity relief ratios. Under the revised metric convention adopted in this paper, these are also gross obligation reduction ratios, since $GOR = LR$ for both procedures.

**Table 4** - Monthly Comparison Year 2021

| Month | # invoices | $D(G_m)$ Invoice amount (€) | Netting GOR=LR (€) | in % | Byppay GOR=LR (€) | in % | Improvement ratio |
|---|---|---|---|---|---|---|---|
| 1 | 13,325 | 1,165,955,548.01 | 221,080,718.13 | 18.96% | 789,612,420 | 67.72% | 357.16% |
| 2 | 12,272 | 1,170,956,240.82 | 219,968,481.95 | 18.79% | 483,494,287 | 41.29% | 219.80% |
| 3 | 13,119 | 1,160,429,687.39 | 259,861,517.69 | 22.39% | 614,118,448 | 52.92% | 236.33% |
| 4 | 11,488 | 1,420,442,106.63 | 241,467,732.07 | 17.00% | 881,397,558 | 62.05% | 365.02% |
| 5 | 10,991 | 1,089,548,131.18 | 202,620,868.35 | 18.60% | 575,433,204 | 52.81% | 284.00% |
| 6 | 11,788 | 1,050,270,469.92 | 259,847,584.37 | 24.74% | 484,172,096 | 46.10% | 186.33% |
| 7 | 11,838 | 1,478,592,075.19 | 251,835,276.97 | 17.03% | 830,188,697 | 56.15% | 329.66% |
| 8 | 10,421 | 1,465,010,612.36 | 495,188,771.71 | 33.80% | 738,344,625 | 50.40% | 149.10% |
| 9 | 10,290 | 1,788,962,425.14 | 335,403,838.31 | 18.75% | 1,045,177,429 | 58.42% | 311.62% |
| 10 | 9,962 | 2,632,324,663.12 | 312,442,963.82 | 11.87% | 1,378,572,850 | 52.37% | 441.22% |
| 11 | 9,018 | 2,549,278,698.08 | 625,313,647.31 | 24.53% | 1,518,656,442 | 59.57% | 242.86% |
| 12 | 8,679 | 2,697,266,071.96 | 703,114,947.92 | 26.07% | 1,256,327,613 | 46.58% | 178.68% |
| **TOTAL** | **133,191** | **19,669,036,729.80** | **4,128,146,348.60** | **20.99%** | **10,595,495,669** | **53.87%** | **256.66%** |

Yearly aggregate improvement: 256.66%
Arithmetic Mean Monthly improvement: 275.15% and standard deviation: 88.14%

## 5.3 Interpreting Structural Differences and Deployment Efficiency

The empirical question is not whether either method is globally optimal. The question is how much reduction each deterministic move set can realize when run to its own fixed point on the same monthly invoice graphs. The cycle restricted benchmark stops when the residual graph is acyclic. Byppay stops when no intermediary has both a positive incoming and a positive outgoing residual edge.

The benchmark results show a consistent advantage for Byppay over Netting, both at the monthly level and in the annual aggregate. This advantage is not merely numerical because it reflects structural differences in the settlement primitives used by the two methods and has direct implications for deployment and risk allocation. The discussion below interprets the results along these three dimensions.

### a) Structural Source of the Performance Gap

The difference in performance between Netting and Byppay follows directly from the settlement primitives used by the two methods. Netting is restricted to directed cycles. It reduces obligations by cancelling the minimum amount on each selected circuit and stops once no positive directed cycle remains in the graph. Byppay, by contrast, applies a local path reduction through an intermediary and therefore does not require obligations to be part of a cycle. It can reduce debt along a path of the form A → B → C by pushing a common amount through the intermediary.

This distinction is important in supply chain invoice graphs, where open chains and tree like structures are common. In such networks, a cycle restricted method can leave substantial obligations untouched even after all admissible cycles have been exhausted. A path enabled method can continue to reduce obligations along open paths, which allows it to capture settlement opportunities that cycle-based methods cannot reach. The benchmark results are consistent with this structural difference: Byppay achieves higher liquidity relief than Netting both in the representative monthly comparison and in the full year aggregate.

The stopping conditions of the two methods also differ. Netting reaches a fixed point when the residual graph is acyclic. Byppay reaches a fixed point when no intermediary has both a positive incoming and a positive outgoing obligation that can support another local reduction. These are not equivalent terminal states. As a result, the path enabled procedure can continue extracting settlement opportunities after the cycle-based procedure has already terminated.

### b) Deployment Implications and Disclosure Requirements

The structural difference between the two methods also has operational implications. Because each Byppay step depends only on an intermediary and its adjacent obligations, it can be executed using local information. This makes the procedure compatible with distributed execution and limits the amount of information that firms must share for a settlement step to be validated.

This is relevant in trade invoice environments, where firms are often unwilling to disclose their full customer and supplier network to a central operator. A local procedure reduces that disclosure burden because only the firms directly involved in the proposed compensation step need to exchange and verify the relevant information. By contrast, cycle-based approaches typically require broader graph visibility in order to identify, rank, and coordinate admissible settlement circuits. In practice, this pushes the process toward centralised data collection and coordination.

The benefit is therefore not only computational. Byppay is also better aligned with selective disclosure requirements that are common in commercial networks. A firm can participate in compensation without exposing the full structure of its relationships, while a shared ledger or similar coordination layer can record validated steps, resulting payment instructions, and audit trails. In this sense, the benchmark results support not only the effectiveness of the method, but also its plausibility as a decentralised clearing procedure.

### c) Counterparty Settlement and Payment Instruction Dynamics

The higher liquidity relief achieved by Byppay comes with an important trade off. In a cycle-based settlement, existing obligations are reduced without changing the identity of the counterparties. By contrast, Byppay changes the obligation structure by replacing part of an indirect chain with a direct payment instruction or promise to pay. This means that the creditor at the end of the path no longer relies on the intermediary for that compensated amount, but instead relies on the original debtor.

In other words, in a standard cycle, the risk profile of the network remains largely unchanged because existing obligations are simply reduced. In Byppay, the transition from $A \rightarrow B \rightarrow C$ to a direct $A \rightarrow C$ instruction introduces new considerations for the firms involved.

**The $A \rightarrow C$ settlement:** When the algorithm identifies an intermediary $B$ with an incoming debt from $A$ $(d(A,B))$ and an outgoing debt to $C$ $(d(B,C))$, it reduces both by $d = min(d(A,B), d(B,C))$:

> *Edge Reduction*: Both $d(A,B)$ and $d(B,C)$ are decreased by the amount $d$ and one edge is reduced
>
> *Instruction Generation*: A settlement, promise to pay instruction, $p2pay(A, C) \mathrel{+}= d$, is recorded.
>
> *Node Status*: This continues until $B$ is directionally separated, meaning it is either strictly a net payer or a net receiver in the remaining network.

**Counterparty Transformation:** A critical technical and managerial implication is the promises transfer from $B$ to $C$, that is the potential exposure reallocation under novation-based despite of no change on maturity of invoices and, therefore, in that debt. The interpretation of this mechanism depends on the implementation layer. Under a strict novation-based implementation, a local path reduction may reallocate effective counterparty exposure because part of the economic claim is redirected from the intermediary toward the upstream debtor. However, that stronger assumption is not required by the model used in this paper. Under the lighter settlement instruction or promise to pay interpretation adopted here, Byppay changes the settlement path without intrinsically replacing the underlying invoice claim. In that case, the benchmark should be read primarily as a comparison of liquidity relief, settlement simplification, and admissible move sets, while any legal transfer of exposure remains contingent on deployment design.

> *Exposure*: Before the step, $C$ held a claim against $B$. After the step, $C$ receives or relies on a settlement instruction from $A$.
>
> *Risk Mitigation*: While this may seem to introduce unknown risk to $C$, the algorithm's *agentifiable* nature allows firms to set constraints. $C$ could potentially whitelist certain debtors $A$ or set limits on the amount of credit risk they are willing to accept from non-direct counterparties.

The mechanism therefore improves liquidity by pushing the settlement promise to pay forward in the path. Before the compensation step, the creditor holds a claim on the intermediary. After the step, the creditor holds a direct claim, or an equivalent payment instruction, on the upstream debtor. This change is central to the economic interpretation of the algorithm. The procedure does not merely simplify the graph. It also reallocates counterparty exposure across firms.

For that reason, practical deployment requires explicit governance around acceptance of redirected obligations. Firms may need prior consent rules, counterparty filters, or whitelists defining which new exposures they are willing to accept. These controls do not weaken the logic of the method but they clarify the conditions under which local path compensation can be implemented in real trade networks. Under such conditions, the evidence in this study indicates that path enabled local compensation can recover substantially more liquidity than cycle restricted netting, while remaining compatible with selective disclosure and decentralised execution.

## 5.4. Complexity Analysis of Byppay for Decentralised Execution

Let $V$ denote the set of firms and $E$ the set of aggregated invoice obligations in the directed graph $G$. Let $n = |V|$ and $m = |E|$. The total running time is obtained by combining two facts: the algorithm executes at most $O(m)$ steps, because each step removes at least one positive edge, and with heap based priority queues each step can be implemented in $O(\log n)$ time. Hence the overall running time is $O(m \cdot \log n)$

Maximum Number of Byppay steps, as in each executed step $A \rightarrow B \rightarrow C$, as the algorithm reduces two adjacent edges by $d = \min(d(A,B), d(B,C))$ because it subtracts the minimum of the two weights, at least one of the two edges is reduced to exactly zero and removed from the graph. The settlement or payment instruction *p2pay(A, C)* is recorded separately and not added back to the active invoice graph $G$. Therefore, the maximum number of executed steps is bounded by the total number of initial positive edges, which is $O(m)$.

Cost per Step (Deterministic Selection) is as follows: The algorithm iterates through nodes $B$ and selects the highest-weight incoming edge from $A$ and highest-weight outgoing edge to $C$.

- *Naive Implementation:* If we scan the adjacency list of $B$ every time, finding the maximum edges takes $O(deg(B))$ time. In the worst-case dense graph scenario, this could lead to $O(m \cdot \Delta)$ where $\Delta$ is the maximum degree of a node, pushing worst-case running time toward $O(m^2)$.

- *Optimized Implementation:* If the graph is implemented using max-priority queues (or heaps) for the incoming and outgoing edges of every node, initializing these structures takes $O(m \log n)$ time. Extracting the maximum edges $d(A,B)$ and $d(B,C)$ takes $O(\log n)$ time. Updating the weight of the non-zeroed edge and re-inserting it takes $O(\log n)$ time.

Using the optimized approach (assuming a priority-queue implementation for edge selection) supports the claim that the local procedure is compatible with decentralised execution, as an $O(m \log n)$ runtime scales highly effectively for sparse supply chain graphs where $m$ is relatively small compared to a fully connected network.

By trading global optimization for a greedy reduction procedure run to exhaustion, Byppay achieves sound, useful scalability. The actual bottleneck in deploying this system is not the algorithm's calculation speed but the network latency of the blockchain, smart contract execution times, and gathering the consents from the companies involved.

# 6 Motivation for Decentralised Operation of Debt Relief Algorithms

Our central question in multilateral debt clearing is not only how much debt a method can reduce, but also whether it can be deployed in a decentralised way at very large scale. In trade invoice networks, this issue is especially important because the firms involved are numerous, heterogeneous, privacy sensitive, and only partially connected. A method may be mathematically effective and still remain difficult to scale if it depends on global visibility of the obligation graph, centralised optimisation, or tightly synchronised batch coordination. For that reason, the relevant comparison is not only between clearing outcomes, but also between the execution models that different algorithms require.

## 6.1. Decentralised developments

From this perspective, blockchain or distributed ledger technology can improve auditability, traceability, and shared state management, but it does not by itself remove the structural limits of the underlying clearing algorithm. If a method requires full graph reconstruction or coordinated circuit selection, placing it on a blockchain does not make it intrinsically decentralised. It only changes how actions are recorded and verified. Similarly, agentic AI can automate discovery, negotiation, validation, and execution of clearing steps, but it cannot overcome a structural dependence on globally visible cycles or computationally expensive optimisation routines. DLT and agentic AI should therefore be understood as execution and coordination layers, not as substitutes for a scalable settlement primitive.

This distinction becomes important when comparing the main approaches in the literature. Exact optimisation methods offer strong theoretical performance, but they assume global graph visibility and become difficult to scale computationally. Heuristic and metaheuristic methods reduce computational cost, but they still rely on centralised search and full network knowledge. Cycle-based services such as Romanian Netting demonstrate that debt clearing can be deployed at national scale, yet they remain dependent on circuit discovery and broad coordination. Privacy preserving decentralised protocols such as **Cycles Protocol** by Buchman, et al. (2024) move further toward distributed execution, but they do so through a heavier coordination stack, combining encrypted intent posting, graph optimisation, and batched atomic settlement. These are meaningful advances, although they still rely on a pooled clearing logic and a coordinated epoch structure.

By contrast, a local path-based method changes the scaling properties of the problem because it changes the settlement primitive itself. Rather than searching for globally coordinated cycles, it applies repeated three-party reductions across local obligations. This matters for decentralisation because a local move can be discovered, proposed, validated, and executed using only neighbourhood information. As a result, the clearing procedure becomes more compatible with distributed execution across a large network of firms without requiring pooled disclosure of the full invoice graph.

From a deployment point of view, this makes local path compensation especially relevant for infrastructures based on DLT or agentic AI. A distributed ledger can provide an auditable record of validated compensation steps, accepted payment instructions, and residual obligations, while autonomous software agents can monitor local invoice positions, identify admissible triples, request consent, verify constraints, and coordinate execution. In such a design, the ledger supplies shared trust and non-repudiation, while the agents supply local intelligence and operational automation. The key advantage is that neither layer requires a central operator to reconstruct the whole graph before clearing can begin.

This is the sense in which decentralisation and scale are linked. The largest scale is unlikely to be achieved by methods that require every participant to reveal its full customer and supplier network to a central hub or to wait for globally coordinated batch optimisation. In fragmented trade networks, scale is more plausibly achieved by methods that minimise disclosure, localise computation, and allow parallel execution across many nodes. The more local the admissible move, the lower the coordination burden and the lower the barrier to participation.

**Table 5** - Comparison of clearing approaches by their compatibility with decentralised execution at scale

| **Method or study** | **Clearing primitive** | **Execution model** | **Decentra-lisation comp-atibility** | **Scalability outlook** | **Main strengths and limits** |
|---|---|---|---|---|---|
| Dynamic programming optimum (Pătcaş, 2011) | Exact optimisation | Central solver | Low | Limited by global computation and full graph visibility | Optimal outcome, but not naturally suited to distributed execution |
| Evolutionary heuristics (Pătcaş and Bartha, 2019) | Metaheuristic search | Central solver | Low to moderate | Better computationally than exact methods, but still centralised | Faster search, but no guarantee of optimality and no local execution logic |
| Romanian Netting (Gavrilă and Popa, 2021) | Directed cycle cancellation | Central service | Low | Proven at national scale, but operationally centralised | Strong real-world deployment record, but depends on cycle discovery and broad disclosure |
| Permissioned consortium netting (CLSNet, Hyperledger pilots) | Netting workflow automation | Consortium DLT | Moderate | Scales within governed networks | Better auditability and access control, but still usually relies on coordinated central or consortium logic |
| Cycles Protocol (Buchman et al., 2024) | Privacy preserving multilateral cycle settlement | Decentralised protocol with batching | High | Strong decentralised design, but with heavier coordination requirements | Strong privacy and verifiability, but protocol complexity and batch cadence may constrain operational simplicity |
| Blockchain invoice finance registries (Battaiola et al., 2019; Mohammadzadeh et al., 2021) | Invoice verification and anti-fraud controls | DLT registry | Moderate | Scales as verification infrastructure, not as clearing logic | Valuable for auditability and fraud reduction, but not designed for network level obligation reduction |
| Byppay | Local three-party path compensation | Peer to peer local execution, optionally supported by DLT for auditability and by software agents for orchestration | High | Strong compatibility with distributed, parallel, and privacy constrained operation | Local settlement rule compatible with selective disclosure and distributed execution, with strong benchmark performance; however, outcomes depend on execution order and on governance rules for redirected obligations. |

This comparison highlights a gap in the state of the art. Existing work provides either strong clearing performance under centralised assumptions, or decentralised architectures with heavier privacy and coordination stacks. What remains underexplored is whether a simple local clearing primitive can support large scale decentralised execution while still achieving strong debt reduction on real invoice networks. That is the gap addressed in this paper.

The motivation for Byppay follows directly from this gap. The contribution of the method does not lie in blockchain infrastructure or agent-based automation as such. Its contribution lies in the settlement primitive: a local path-based compensation rule that can reduce obligations using only neighbourhood information. This matters because a genuinely local rule is structurally more compatible with decentralised execution than methods that depend on global graph visibility or coordinated cycle enumeration. In that setting, DLT can provide an auditable shared record of validated compensation steps, accepted payment instructions, and residual obligations, while agentic AI can support the operational layer by monitoring local positions, proposing admissible triples, checking constraints, requesting consent, and coordinating execution. These technologies therefore strengthen deployability and scale, but they are not the source of the algorithmic contribution itself.

In other words, DLT and agentic AI are best understood here as execution and coordination layers built around the clearing rule, rather than as part of the rule itself.

The research question is therefore not only whether a path enabled local method can clear more debt than a cycle restricted baseline, but also whether it offers a better execution model for decentralised scaling. This leads to the following question:

**RQ. To what extent can a decentralisable local debt compensation algorithm outperform established and centralised cycle-based netting methods while remaining more compatible with large scale distributed execution?**

This study contributes in this direction. As it formalises a local path-based compensation rule for invoice clearing that is compatible with decentralised execution, and then it benchmarks that rule directly against the Romanian Netting baseline on identical monthly and annual invoice graphs, we proceed to interpret the results in deployment terms, showing why local settlement primitives are more naturally aligned with selective disclosure, DLT based auditability, and agent-based automation than methods that depend on broader graph visibility and coordinated cycle enumeration.

Taken together, the motivation for this study is practical as much as analytical. If multilateral invoice clearing is to operate across very large, fragmented, and privacy sensitive supply chains, then the relevant question is not only how much debt can be cancelled, but what kind of algorithm can realistically be executed in a distributed environment. This paper argues that scale is more likely to be achieved when the settlement primitive itself is local, and it evaluates that proposition empirically on national scale invoice data.

## 6.2. Discussion on its Implementation Possibilities on a Blockchain and Deployment Implications

A permissioned blockchain provides a plausible execution layer for Byppay because it can support automation, auditability, and controlled information sharing without altering the underlying settlement primitive. In the architecture considered here, the ledger records validated settlement instructions or promises to pay rather than necessarily novated legal claims. This distinction is important. The contribution of Byppay lies in the local compensation rule itself, whereas the blockchain serves as an execution, coordination, and traceability layer. Whether validated instructions are subsequently mapped into legal novation depends on the governance framework of the deployment.

The suitability of this architecture follows from the locality of the Byppay move. Each compensation step involves only a directed 2 path, A → B → C, and therefore requires verification of only the two adjacent obligations affected by the reduction. A smart contract can implement this logic by verifying the relevant balances, reducing the obligations on (A, B) and (B, C) by the matched amount, and recording the resulting settlement instruction from A to C. Unlike cycle-based procedures, this design does not require global reconstruction of the obligation graph or centralised discovery of admissible clearing structures.

Locality also reduces the coordination burden associated with decentralised execution. The main operational risk is concurrent processing of the same invoice edge by more than one intermediary. In a permissioned consortium setting, this can be managed through lightweight atomicity controls, such as temporary locks on affected edges or a consensus mechanism that serialises updates for the small set of obligations involved in a proposed step. Because only a limited neighbourhood of the graph is touched by each execution, these safeguards can be implemented without the heavy coordination requirements associated with globally synchronised clearing procedures.

A further implication concerns information disclosure. Since the validity of a Byppay step depends only on the existence and weights of two adjacent obligations, firms need not reveal their full supplier and customer network to a central operator. The architecture is therefore compatible with selective disclosure. If stronger privacy guarantees are required, cryptographic verification techniques, including zero knowledge proofs, could be incorporated to confirm the validity of the matched reduction without exposing broader balance information to the wider network.

Finally, the ledger can function as an auditable bulletin board for downstream legal and operational processes. Timestamped and signed records of validated settlement instructions provide a verifiable trail for compliance, accounting, and dispute resolution. In that sense, the blockchain layer strengthens enforceability and reduces reconciliation costs, while the algorithm remains a local settlement procedure. The broader implication is that Byppay combines strong clearing performance with an execution model that is more compatible with privacy sensitive and distributed trade networks than conventional centralised netting systems.

## 7. Conclusions and Future Work

This paper shows that path enabled compensation is structurally more effective than cycle restricted netting in invoice-based trade credit networks. The contribution is methodological and empirical: it shows that when a path enabled local move set is formalized for invoice networks and evaluated on a national invoice corpus, it reaches a substantially larger reduction fixed point than a cycle restricted benchmark while remaining more compatible with distributed execution. Using the 2021 IMI corpus in Romania, we find that Byppay consistently outperforms the Netting baseline because it acts on a broader set of settlement opportunities. Cycle based netting can reduce obligations only when debt lies on directed circuits, and it stops once the residual graph becomes acyclic, even when substantial open chains remain. Byppay removes this restriction by propagating reductions along local paths of the form A → B → C, thereby capturing debt relief opportunities that cycle only mechanisms leave untouched. In supply chain graphs, where open paths are pervasive, this advantage is structural. The local path primitive was known in the broader graph simplification literature, but this paper contributes by adapting it to invoice networks, benchmarking it reproducibly against a cycle-based reference, and framing it for realistic local execution.

A key clarification concerns measurement. Earlier formulations risked conflating invoice layer compression with liquidity relief. In the formulation adopted here, gross obligation reduction is defined on the combined post settlement payable state, so that $GOR = LR$ for both Netting and Byppay. Invoice layer compression is retained only as a secondary structural metric. This strengthens the comparability of the benchmark and keeps the economic interpretation clear.

The contribution of the paper therefore takes an existing debt simplification primitive to a novel application of a known local path reduction rule to invoice backed trade networks, its formalization as a deployable protocol under selective disclosure and consent constraints, and its benchmarking at national scale against a real operational system. That combination shifts the discussion from abstract settlement logic to an empirically grounded protocol that is analytically interpretable and operationally scalable.

A second contribution is architectural. Because Byppay relies on a local move set that preserves firms' net positions and runs until no admissible path reduction remains, it offers a practical advantage beyond aggregate performance. Each firm needs visibility only over its own receivables and payables, with coordination limited to direct counterparties. By contrast, cycle-based procedures usually require broader graph visibility to identify and coordinate eligible circuits. Byppay is therefore better aligned with distributed execution, agent-based orchestration, and privacy constrained environments.

The main deployment trade-off is also clear. Under the promise to pay interpretation adopted here, Byppay should be understood primarily as a settlement layer that redirects execution while leaving the underlying invoice claims legally intact until discharge. Read in that way, the comparison with cycle restricted netting turns on the breadth of the admissible move set and the resulting liquidity relief, rather than on any necessary redefinition of the underlying claims. Path propagation improves liquidity by thinning the network of obligations, but it can also reallocate credit exposure. When a Byppay step replaces $A \rightarrow B$ and $B \rightarrow C$ with a direct settlement from A to C, creditor C exchanges a claim on B for a claim on A. That risk transfer does not weaken the empirical result, but it does define the governance requirements of any production deployment. Practical implementations should therefore allow consent rules, whitelists, and exposure limits before firms accept redirected payment instructions. Future work should relax the full participation assumption and test how partial consent, credit filters, and exposure caps affect realised relief.

The empirical grounding of the study is one of its main strengths. To our knowledge, this is among the largest comparative benchmarks of multilateral debt compensation on a real invoice corpus. The evidence shows not only that Byppay improves aggregate relief, but also why it does so: it matches the topology of trade credit networks more closely than cycle restricted systems.

Future work should develop the paper in two directions. The first is robustness analysis. Alternative deterministic tie breaking rules, different edge orderings, and multiple valid execution schedules should be tested to determine how stable the performance advantage remains across admissible implementations. Longer path reductions, including general n edge Byppay extensions, and hybrid procedures combining cycle settlement with path propagation may unlock additional relief. The second direction is deployment realism. Future studies should examine partial consent, dynamic credit risk filters, settlement costs, payment rail constraints, timing effects, credit ceilings, invoice age-based prioritisation, multi-currency compensation, and incentive layers that reward firms for facilitating settlement. These are not weaknesses of the present paper, but natural next steps once benchmark superiority has been established under a transparent and reproducible benchmark design.

The results can be read in two layers. Temporal prevalence means that Byppay can create usable liquidity earlier, before a circuit exists and before due dates are missed. Topological prevalence means that Byppay can act on a larger structural opportunity set because invoice networks typically contain many more local intermediary paths than closed debt circuits. Together, these results show that paths precede circuits in time, outnumber them in feed forward trade networks, and therefore support a larger space of admissible reductions. In that sense, the paper's strongest contribution is not merely the use of a path shortening move, but its integration into a selective disclosure protocol for invoice networks, supported by large scale empirical validation against a real cycle-based baseline.

**"URLs and whitepapers"**